\documentclass[preprintnumbers, prd, showpacs, onecolumn, floatfix, preprintnumbers, letterpaper, amsmath, amssymb, superscriptaddress]{revtex4}

\usepackage{amssymb, amsmath, bm, dcolumn, epsf, graphicx, latexsym, mathbbol, slashed}

\usepackage{color}

\newcommand{\be}{\begin{equation}}
\newcommand{\ee}{\end{equation}}
\newcommand{\bea}{\begin{eqnarray}}
\newcommand{\eea}{\end{eqnarray}}
\newcommand{\barr}{\begin{array}}
\newcommand{\earr}{\end{array}}

\begin{document}

\title{Hint of relic gravitational waves in the {Planck} and {WMAP} data}

\author{Wen Zhao}
\affiliation{Key Laboratory for Researches in Galaxies and Cosmology,
Department of Astronomy, University of Science and Technology of China, Hefei, Anhui,
230026, China}

\author{Cheng Cheng}
\affiliation{State Key Laboratory of Theoretical Physics,
Institute of Theory Physics, Chinese Academy of Science, Beijing
100190, China}
\affiliation{University of the Chinese Academy of Sciences, Beijing 100190, China}

\author{Qing-Guo Huang}
\affiliation{State Key Laboratory of Theoretical Physics,
Institute of Theory Physics, Chinese Academy of Science, Beijing
100190, China}


\begin{abstract}
{Relic gravitational waves (RGWs) leave well-understood imprints
on the anisotropies in the temperature and polarization of cosmic
microwave background (CMB) radiation. In the TT and TE information
channels, which have been well observed by WMAP and Planck
missions, RGWs compete with density perturbations mainly at low
multipoles. It is dangerous to include high-multipole CMB data in
the search for gravitational waves, as the spectral indices may
not be constants. In this paper, we repeat our previous work
[W.Zhao \& L.P.Grishchuk, Phys.Rev.D {\bf 82}, 123008 (2010)] by
utilizing the Planck TT and WMAP TE data in the low-multipole
range $\ell\le100$. We find that our previous result is confirmed
{with higher confidence}. The constraint on the tensor-to-scalar
ratio from Planck TT and WMAP TE data is $r\in [0.06,~0.60]$ (68\%
C.L.) with the maximum likelihood at around $r\sim 0.2$. Correspondingly, the
spectral index at the pivot wavenumber $k_*=0.002$Mpc$^{-1}$ is
$n_s=1.13^{+0.07}_{-0.08}$, which is
 larger than 1 at more than $1\sigma$ level. So, we conclude that the new released
CMB data indicate a stronger hint for the RGWs with the amplitude
$r\sim 0.2$, which is hopeful to be confirmed by the imminent
BICEP and Planck polarization data. }
\end{abstract}

\maketitle

\section{Introduction}

The relic (primordial) gravitational waves generated in the early
Universe is a basic prediction in the modern cosmology, which
depends only on the validity of General Relativity and Quantum
Mechanics \cite{grishchuk1974,starobinsky1980}. The relic
gravitational waves (RGWs) leave the imprints in all the cosmic
microwave background (CMB) radiation anisotropy power spectra,
including the TT, TE, EE and BB. In the near future, these provide
the unique way to detect it in the observations. If the amplitude
of the RGWs is large, (i.e. the tensor-to-scalar ratio $r>0.1$),
the CMB TT and TE information channels can dominate the detection,
since the amplitudes of these spectra generated by RGWs are much
larger than those of EE and BB \cite{turner1994,zhao2009a}.
However, if $r<0.1$, these channels become useless due to the
cosmic variance, and the detection can only be done through the
B-mode polarization \cite{zaldarriaga1997,kamionkowski1997}.

In the era before the release of the Planck polarization data, the
detection (or constraint) of RGWs mainly depends on the CMB TT and
TE channels, which has been done by many groups, including the
{WMAP} and {Planck} teams. It is well known that the TT and TE
power spectra generated by RGWs are significant only in the large
scales, i.e. the low multipoles $\ell\lesssim 100$. However, in
the previous analyses, nearly all the groups utilized the full CMB
data till to the very high multipoles ($\ell_{\max}\sim 1200$ for
WMAP and $\ell_{\max}\sim 2500$ for Planck), and assumed density
perturbations with a constant or a running spectral index. This
can easily overlook the contribution of RGWs, due to the
degeneracies among various cosmological parameters (in particular,
the degeneracy between $r$ and $n_s$).

In 2006, Basksran, Grishchuk and Polnarev noticed that the WMAP TE
data are systematically smaller than the predictions of the
best-fit cosmological model \cite{baskaran2006,grishchuk2007},
where the RGWs are absent, and argued that this might hint the
existence of RGWs. In 2009, for the first time, one of us (W.Zhao)
with Baskaran and Grishchuk carefully analyzed the three-year WMAP
TE data in the low multipoles $\ell\le 100$, and gotten the
constraints on the quadrupole ratio $R=0.149^{+0.247}_{-0.149}$
(note that the tensor-to-scalar ratio is $r\simeq 2R$)
\cite{zbg2009a}. In addition, we have extended this analysis to
the five-year and seven-year WMAP TT and TE data in the low
multipoles $\ell\le 100$, and found that the indication of RGWs
were stabilized: five-year data give $R=0.266\pm 0.171$
\cite{zbg2009b}, and seven-year data yield
$R=0.273^{+0.185}_{-0.156}$ \cite{zbg2010}. In these analyses, we
have adopted an approximate effective noises and the likelihood
functions for the WMAP data, which are based on the exact Wishart
distribution for the full-sky observables. However, these
approximations were questioned by some authors (see for instance
\cite{blame}). To clarify it, in paper \cite{zg2010}, we adopted
the commonly used CosmoMC numerical package to repeat the WMAP7
analysis. We found the maximize likelihood (ML) values are
$r=0.285$ and $n_s=1.052$, and one-dimensional (1d) marginalized
likelihood gives the constraints: $r=0.20^{+0.25}_{-0.20}$ and
$n_s=1.064^{+0.058}_{-0.059}$. The CosmoMC approach reduced the
confidence of the indications from approximately 2$\sigma$ level
to approximately 1$\sigma$ level, but the indications do not
disappear altogether.

Recently, Planck team released their CMB TT data, and shown some
differences in the low multipoles compared with the WMAP data
\cite{planck2013,planck2013_2}. In this paper, we shall repeat the
analysis in \cite{zg2010} based on the combination of Planck TT
data and nine-year WMAP TE data \cite{wmap9}, and investigate the
hint of RGWs in these new data, where the public CosmoMC numerical
package is used for the data analysis. As anticipated, we found
that the new data favor the gravitational waves with {$r\sim
0.2$}, and a blue tilted spectrum of density perturbation with
{$n_s\sim1.08$}. So, the new data stabilize what we found in the
previous work \cite{zg2010}.

\section{Data analysis method and the parameter constraints}

Relic gravitational waves compete with density perturbations in
generating CMB temperature and polarization anisotropies at low
multipoles $\ell\lesssim100$. Therefore we focus on the Planck TT
data and WMAP9 TE data at $\ell\le100$. Limited by the data number
in our analysis, it is impossible to determinate all the
cosmological parameters together. Similar to our previous works
\cite{zbg2009a,zbg2009b,zbg2010,zg2010}, we fix the background
parameters at their best-fit values in the $\Lambda$CDM model
\cite{planck2013_2}: $\Omega_bh^2=0.022032$,
$\Omega_ch^2=0.12038$, $100\theta_{\rm MC}=1.04119$,
$\tau=0.0925$. The free parameters subject to evaluation by the
data analysis are the parameters: the amplitude and the spectral
index of density perturbations $\ln(10^{10}A_s)$, $n_s$, and the
tensor-to-scalar ratio $r$. Note that, throughout this paper, we
shall adopt the spectral amplitudes and the spectral indices at
the pivot wavenumber $k_*=0.002$Mpc$^{-1}$.

The value of $n_t$, the spectral index of RGWs, is very difficult
to be determined by the current data. However, most inflationary
models predict the nearly scale-invariant spectrum with
$n_t\approx0$. In {\it Case I}, the inflationary consistency
relation $n_t=-r/8$, which is valid for the single-field slow-roll
inflationary models, is adopted. By running the public CosmoMC
code, we derive the following 3d ML values of the perturbation
parameters:
 \begin{equation}
 r=0.20, ~~~~n_s=1.08, ~~~~\ln(10^{10}A_s)=3.02.
 \end{equation}
The marginalized 1d results for these parameter (see also Table I)
are
 \begin{equation}
 r\in[0.06,~0.60], ~~~~n_s=1.13^{+0.07}_{-0.08}, ~~~~\ln(10^{10}A_s)=2.94^{+0.14}_{-0.12}.
 \end{equation}
Note that, throughout this paper, we quote the mean values of 1d
likelihood functions, and/or the uncertainties refer to the $68\%$
confidence intervals. The 1d likelihood functions for $n_s$ and
$r$ are plotted in Fig. 1. In the upper panel, the black curve
shows a significant peak around $r\sim 0.3$, and $r=0$ is excluded
at more than $1\sigma$ level. Correspondingly, from the low panel
we see that the spectral index $n_s$ is larger than $1$ at more
than $1\sigma$ level. Interesting enough, we find that all these
new results are compatible with the previous ones in
\cite{zg2010}, and also the indications of RGWs becomes higher for
the better data.

In addition, we also consider other two cases: {\it Case II} with
$n_t=0$ and {\it Case III} with $n_t=n_s-1$. The former case is
nearly kept for the most inflationary models. The latter case with
$n_t=n_s-1$ is insisted by some authors
\cite{grishchuk1974,grishchuk2007}, which is valid if the
expansion history of inflation can be approximated as a power-law
form of the conformal time. The results for these two cases are
presented in Table 1 and in Fig. 1 where the red and blue curves
correspond to {\it Case II} and {\it Case III}, respectively. From
Table 1 and Fig. 1, we see that the results in both {\it Case II}
and {\it Case III} are quite similar to those in {\it Case I}. We
conclude the assumption on $n_t$ cannot significantly influence
our results.

Before the end of this section, we need to mention that there is
some awkwardness in the choice of the background parameters, as
these are the parameters derived by Planck \cite{planck2013_2}. Actually the choice of these background parameters does not significantly affect the TT and TE spectra at low multipoles \cite{Challinor:2004bd,challinor2005}. As
we know from our previous experiences \cite{zbg2010,zg2010}, the
background parameters, if changed not too much, do not
significantly affect the results.
In order to make sure of it, we also fixed the background parameters to be WMAP best-fit values, and obtained similar results.
Nevertheless, for the safety, we
shall explicitly explore the issue of background parameters in a
separate work \cite{new_work}.

\begin{table}
\caption{Results for $n_s$ and $r$ in Cases I-III}
\begin{center}
\label{tab0}
\begin{tabular}{ l  c  c  c  c}
   \hline
   \hline
   &  \multicolumn{2}{c}{~~~~~Maximum likelihood~~~~~~} & \multicolumn{2}{c}{~~~~~1-d likelihood~~~~~~} \\
   \hline
   ~~~   & ~~~~~~~~~$n_s$~~ & ~~~~~$r$~~~~~ & ~~~$n_s$~~~ & ~~$r$ (68\% C.L.)~~          \\
   Case I ~~($n_t=-r/8$)~~   & ~~~~~~~1.08~~ & ~~~~~0.20~~~~~ & $~~1.13^{+0.07}_{-0.08}~~$ & ~~[0.06,~0.60]~~          \\
   Case II ~($n_t=0$)~~~   & ~~~~~~~1.09~~ & ~~~~~0.25~~~~~ & $~~1.12^{+0.06}_{-0.08}~~$ & ~~[0.00,~0.52]~~          \\
   Case III ($n_t=n_s-1$) & ~~~~~~~1.07~~ & ~~~~~0.24~~~~~ & $~~1.11^{+0.05}_{-0.07}~~$ & ~~[0.05,~0.51]~~          \\
   \hline
\end{tabular}
\end{center}
\end{table}

\begin{figure}
\begin{center}
\includegraphics[width=9cm]{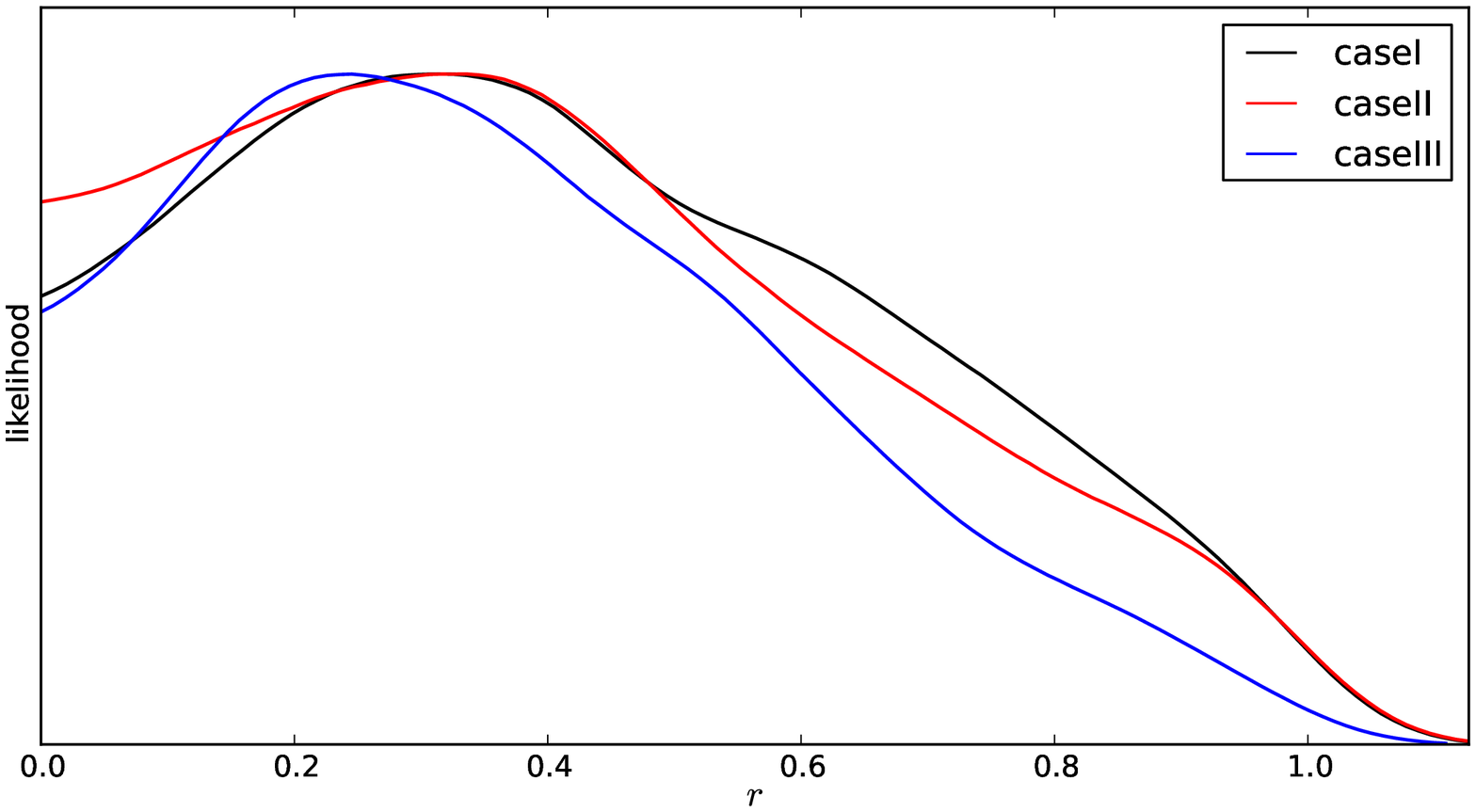}
\includegraphics[width=9cm]{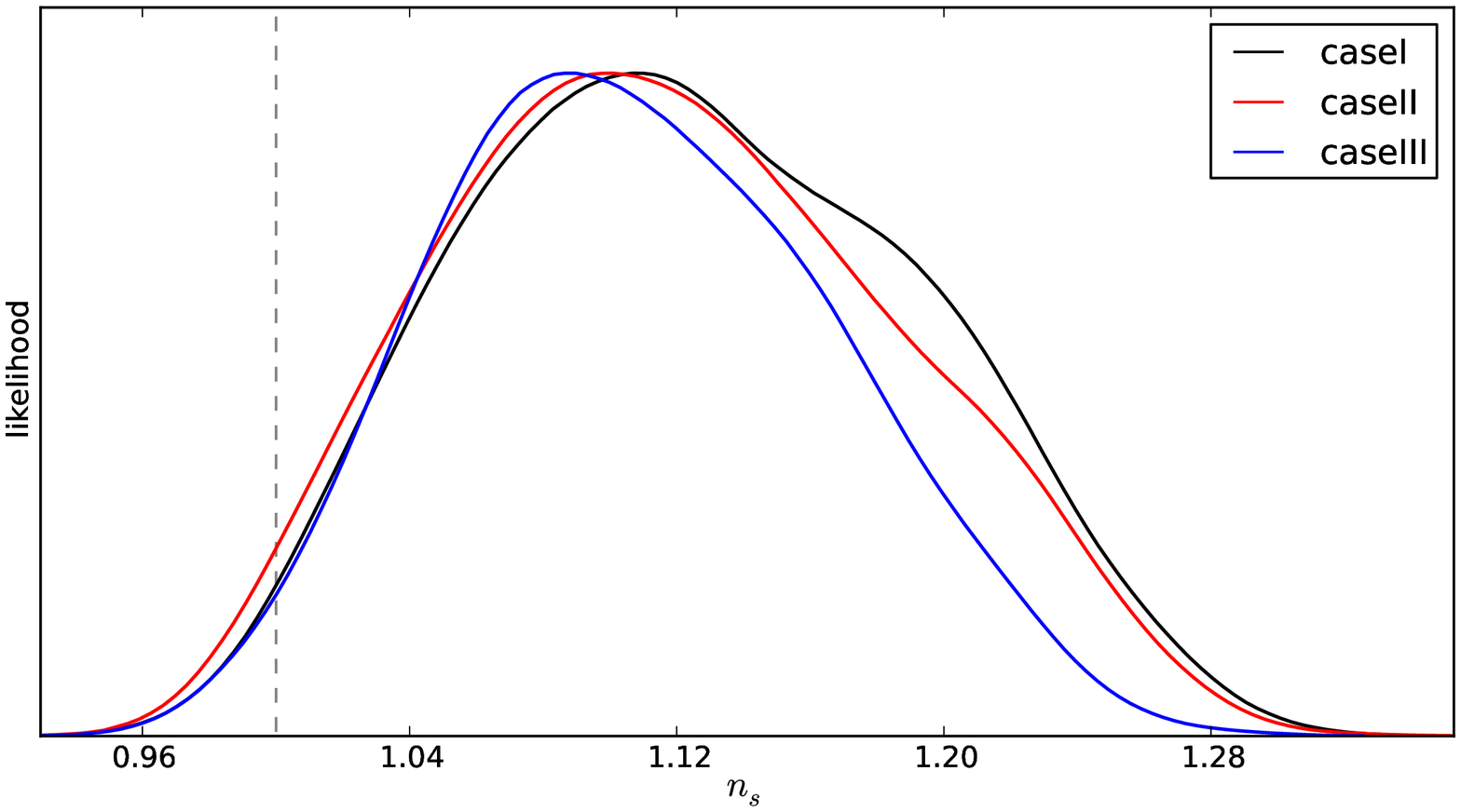}
\caption{\label{fig1} One-dimensional likelihood functions for $r$
(upper panel) and $n_s$ (lower panel) in Cases I-III. }
\end{center}
\end{figure}

\section{Conclusions \label{sec4}}

Relic gravitational waves provide the unique antenna to study the
expansion history of the very early Universe. The detection of
RGWs through their imprints in the CMB temperature and
polarizations anisotropies is the only possibility in the near
future, which has also been considered as one of the key tasks for
the current and future CMB observations. In the well observed CMB
TT and TE information channels, RGWs compete with density
perturbations only in the low-multipole range. So, it is sensible
to utilize only the low-multipole data in the search of RGWs,
which is helpful to keep away from the unwarranted assumptions
about density perturbations, and avoid the oversight of RGWs in
the data analysis. In this paper, we repeated our previous
analysis in \cite{zg2010} by considering the low-multipole Planck
TT data, as well as the nine-year WMAP TE data. We found that, the
new data give the constraint $r\in[0.06,~0.60]$ at $68\%$
confidence level, which deviates from zero at more than $1\sigma$
confidence level. Meanwhile, the data favor a blue tilted spectra
of primordial density perturbations with the spectral index
$n_s=1.13^{+0.07}_{-0.08}$ in the large scale. All these are
consistent with what we found in \cite{zg2010}. We hope the
forthcoming CMB polarization data of BICEP experiment and Planck
mission could confirm our expectations.

\vspace{5mm}
\noindent {\bf Note:} in the same day BICEP \cite{bicep2} released its data which indicates a discovery of the primordial gravitational waves with $r=0.20_{-0.05}^{+0.07}$ and $r=0$ disfavored at $7.0\sigma$.

\vspace{5mm}
{\it Acknowledgments:} WZ would like to dedicate this article to
his friend Leonid Petrovich Grishchuk, who passed away in 13th,
September 2012. We acknowledge the use of Planck Legacy Archive, ITP and Lenovo Shenteng 7000 supercomputer in the Supercomputing Center of CAS for providing computing resources. WZ is supported by
project 973 under Grant No.2012CB821804, by NSFC No.11173021,
11322324 and project of KIP of CAS. QGH is supported by NSFC
No.10821504, 11322545, 11335012 and project of KIP of CAS.

\appendix

\end{document}